# Axially open nonradiative structures: an example of single-mode resonator based on the sample holder

G. Annino[@], M. Cassettari, M. Martinelli
*Istituto per i Processi Chimico-Fisici, Area della Ricerca CNR, via G. Moruzzi 1, 56124 Pisa (Italy)*
(February 1, 2005)

**Abstract**

The concept of nonradiative dielectric resonator is generalized in order to include axially open configurations having rotational invariance. The resulting additional nonradiative conditions are established for the different resonance modes on the basis of their azimuthal modal index. An approximate chart of the allowed dielectric and geometrical parameters for the $TE_{011}$ mode is given. A practical realization of the proposed device based on commercial fused quartz tubes is demonstrated at millimeter wavelengths, together with simple excitation and tuning mechanisms. The observed resonances are characterized in their basic parameters, as well as in the field distribution by means of a finite element method. The predictions of the theoretical analysis are well confirmed, both in the general behaviour and in the expected quality factors. The resulting device, in which the sample holder acts itself as single-mode resonating element, combines an extreme ease of realization with state-of-the-art performances. The general benefits of the proposed open single-mode resonators are finally discussed.

_e-mail address: geannino@ipcf.cnr.it_

# 1. Introduction

The search for efficient millimeter-wave components as waveguides, directional couplers, mirrors, gratings, non-reciprocal elements (circulators, faraday rotators) and resonators have been stimulated by the recent development of versatile solid-state sources. Among the above components, the resonators play a fundamental role in each spectroscopic application, due to the increase of the electromagnetic energy density on the sample and to an effective decoupling of the field distribution from the employed propagation circuit. At millimeter wavelengths the commonly employed resonant devices can be divided in overmoded resonators, like Fabry-Perot cavities [1, 2] and whispering gallery mode dielectric resonators [3], and single-mode ones, as the well-known $TE_{011}$ cylindrical cavity [4]. The single-mode resonators are accredited of higher power-to-field conversion factors, due to their minimal active volume. In turn they should guarantee the highest absolute sensitivity in any application where the intensity of the electric or magnetic field on the sample is the key parameter. Specific efforts have been then devoted to the development of efficient millimeter-wave single-mode resonators [5-7]. The most problematic issues to face in this case are given by the size of the resonator, which is of the order of the employed wavelength, and by the excitation configuration. The commonly proposed solutions, typically borrowed from the microwave technology, are indeed rather difficult to be realized when the wavelength approaches to the millimeter. As an example, the typical size of the coupling hole of a $TE_{011}$ cavity working at 100 GHz is of about 0.8 mm. The corresponding thickness of the cavity wall around this hole must be typically less than 0.05 mm [5]. The common requirement of sample manipulation inside the cavity or of its additional irradiation can complicate further on the design and the realization of the cavity, leading to performances severely reduced in comparison to the ideal ones. An important application where the above aspects are of particular relevance is given by Electron Paramagnetic Resonance (EPR) spectroscopy, which often requires the rotation of the sample and the use of open cavities, in order to allow a proper static magnetic field modulation as well as radiofrequency or optical excitation [6, 8, 9]. A relevant EPR research activity is in particular addressed towards high magnetic fields, where the working wavelengths approach and cross the submillimeter borderline [1, 9-12]. For a proper development of millimeter and submillimeter wave single-mode resonators a novel approach seems then mandatory. A recent proposal of single-mode dielectric resonators specifically designed for millimeter-wave applications is reported in Refs. [13, 14]. The room temperature state-of-the-art conversion efficiency has been obtained in a simple structure by using a partially open NonRadiative (NR) configuration. The same principle has been extended to metallic resonators, allowing the realization of a widely open $TE_{011}$ cavity [15]. In this reference the possibility of open resonators which combine the benefits of both metallic cavities and dielectric resonators was anticipated. The present paper is a further step along this research line, in which the concept of NR resonator is generalized in order to include axially open configurations. In particular, the analysis will focus on single-mode resonances in arbitrarily long dielectric tubes partially shielded by conducting mirrors. In these structures the active region of the resonator can be given by a section of the sample holder itself. The final aim of the paper is the demonstration of open single-mode resonators having a peculiar ease of realization combined to state-of-the-art performances at millimeter wavelengths. The proposed solutions will be analyzed having in mind possible applications to millimeter-wave EPR spectroscopy. Accordingly, the performances of the resonators will be evaluated in terms of their magnetic field conversion factor.

The plan of the paper is the following. Sect. 2 will discuss the physical background underlying the realization of a proper axially open NR resonator. The global NR conditions will be established; the related chart of allowed dielectric and geometrical parameters will be given for the $TE_{011}$ mode. In Sect. 3 a practical realization of the proposed device, based on a commercial fused quartz tube, will be investigated at millimeter wavelengths. Possible tuning mechanisms will be demonstrated as well. The expected field distributions, calculated by using a finite element numerical method, will be presented in Sect. 4. Finally, Sect. 5 will be dedicated to the analysis of the experimental and computational results, and of their implications.

## 2. General aspects

The peculiarity of a NR device is given by its partially open structure, where a region of allowed propagation is surrounded by a (partially open) region of forbidden propagation. A simple application of this principle is given by the NR dielectric resonator [13], which indeed arose as a logical development of the firstly proposed NR waveguide [16, 17]. Figs. 1a and 1b show a cylindrical version of NR resonator, in which a small central hole is included in order to contain the sample. Here the metallic shielding is given by two plane and parallel mirrors, whose distance is indicated with $l$. When the diameter of the dielectric disc is of the order of $l$, the working condition of the resonator can be written as

$$l < \frac{\lambda_0}{2} < l \cdot \sqrt{\varepsilon}, \qquad (1)$$

where $\lambda_0$ is the wavelength in vacuum corresponding to the resonance frequency and $\varepsilon$ the permittivity of the dielectric region. Under the above conditions the central region can contain a resonance mode. The surrounding propagation can be only due to the cutoff-less TEM mode, being the parallel-plates TE and TM modes below their cutoff. Any resonance mode having negligible projection on the TEM mode is thus expected unaffected by irradiation losses, provided that the extension of the mirrors is wide enough. This conclusion was experimentally verified on the $HE_{111}$, $TE_{011}$ and $TM_{011}$ modes of cylindrical NR dielectric resonators in Refs. [13, 14]. The presence of nonradiating modes can be argued for more general configurations on the basis of rigorous symmetry considerations. Indeed, transverse-electric modes with azimuthal invariance, hereafter referred to as $TE_0$ modes, are compatible with any configuration having rotational symmetry, as discussed in [15]. The $TE_0$ modes don't share any field component with the parallel-plates TEM mode (characterized only by axial electric field and azimuthal magnetic field), so at least a nonradiating mode family is expected for any configuration fulfilling the above symmetry property. A simple excitation configuration of these resonators can be obtained exploiting their nonradiative character, as discussed in Ref. [13]. A radiation incident on the NR structure is indeed totally reflected when its polarization is parallel to the conducting mirrors, provided that the employed wavelength satisfies the NR condition. Only an evanescent field can extend inside the mirrors; this field can transfer energy to the resonator when overlapped with the field of the resonance mode. Once the working frequency and the distance between the mirrors are fixed, the basic parameter that imposes the level of coupling is given by the distance between the incoming radiation and the dielectric region. This parameter can be changed for instance by moving the NR device orthogonally to the incoming radiation, as indicated by the dashed line of Fig. 1a [13, 14].

The basic NR configuration reported in Figs. 1a and 1b can be modified as shown in Fig. 1c, in which the central dielectric disc is replaced by a nominally infinite dielectric tube. The rotational invariance ensures the confinement of the $TE_{011}$ mode along the planar aperture, according to the above NR analysis. A planar confinement is expected for the $HE_{111}$ and $TM_{011}$ modes as well, being the general conditions of symmetry unchanged. The new structure can behave as a proper resonator when the propagation along the dielectric tube is prevented. The analysis of this propagation can be based again on general considerations, in which the symmetry of the structure plays a central role. The resonator can be considered as composed by a central dielectric region in contact with two metallic waveguides partially filled with a dielectric tube. Following the arguing developed for the NR configuration, a resonance mode remains confined in the central region when its projection on the modes propagating along the partially filled metallic waveguides vanishes. The modes propagating along these waveguides can be identified by a pair of indices (*n*,*m*), namely an azimuthal index *n* and a radial index *m*, due to the axial symmetry of the structure. The symmetry ensures again that only modes having the same azimuthal index can couple to each other [18]. As a consequence, the analysis of the confinement properties will involve only resonant modes of the central region and propagating modes in the adjacent waveguides having the same index *n*. In the following the different number of modal indices will discriminate between resonant modes and propagating modes.

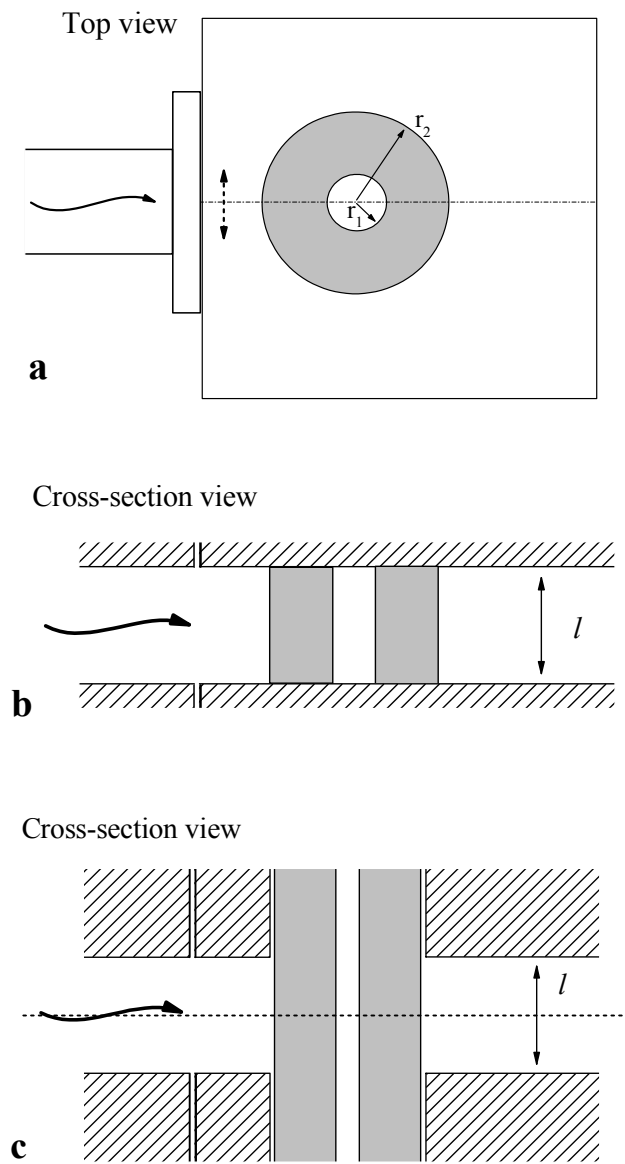

FIG. 1. **a** Top view of a cylindrical NR dielectric resonator with inner radius $r_1$ and outer radius $r_2$. The excitation waveguide is shown. **b** Cross-section view of a planar mirrors NR dielectric resonator. The distance between the mirrors is indicated with *l*. **c** Cross-section view of a possible axially open NR dielectric resonator. The shaded line represents the plane of symmetry of the structure.

The most important mode of any NR structure is given, as discussed above, by the $TE_{011}$ mode. The azimuthally invariant counterpart of the metallic waveguide, assumed here homogeneously filled with a dielectric material, is given by the $TE_{0m}$ family [4, 19]. The $TE_{011}$ mode doesn't share indeed any field component with the other family of modes having vanishing azimuthal index, namely the $TM_{0m}$ family. The cutoff frequencies of the $TE_{0m}$ modes are given by $\nu_{TE_{0m},cutoff} = \frac{u'_{0,m} \cdot c}{2\pi r_2 \cdot \sqrt{\varepsilon}}$, where $c$ is the velocity of light in vacuum, $\varepsilon$ the permittivity of the dielectric region, $r_2$ the radius of the waveguide, and $u'_{0,m}$ the nonvanishing roots of the derivative of the first order Bessel function $J_0$ [20]. The lowest cutoff frequency corresponds to the $TE_{01}$ mode and is given by $\nu_{TE_{01},cutoff} = \frac{3.8317 \cdot c}{2\pi r_2 \cdot \sqrt{\varepsilon}}$. The axial confinement of the $TE_{011}$ mode is then guaranteed provided that $\nu_{TE_{011}} < \nu_{TE_{01},cutoff}$. For a given working frequency the condition of axial confinement can be expressed in terms of the radius of the waveguides as

$$r_{2,TE_{011}} < \frac{3.8317 \cdot \lambda_0}{2\pi \cdot \sqrt{\varepsilon}}. \qquad (2)$$

In order to verify the compatibility of this constraint with Eq. (1), we can go back to the basic structure of Fig. 1b. The meaning of Eq. (2) can be first investigated on this test configuration. The differences with a true axially open resonator will be discussed later on.

In order to model a realistic NR resonator, the mirrors will be assumed made of aluminium, whose resistivity is equal to 2.8 μΩcm. The loss factor $\tan\delta$ of the dielectric region will be assumed equal to $6.7 \cdot 10^{-4}$; its permittivity will be varied from 2 to 11. The resonance frequency of the $TE_{011}$ mode will be fixed to 91.11 GHz. The reason of these somehow arbitrary choices will be cleared up in the experimental section. The dielectric disc will be assumed without central hole for sake of simplicity. By using the above parameters, the resonance frequency, the field distribution and the merit factor $Q_0$ of any mode of the resonator of Fig. 1b can be calculated following the approach of Ref. [13]. In particular, the axial magnetic field conversion factor $B_z$ can be calculated for any allowed pair ($l$, $r_2$) imposed by resonance frequency and dielectric permittivity. The curves obtained for the $TE_{011}$ mode are reported in Fig. 2. Here $B_z$ is reported versus the normalized aspect ratio, defined as $\frac{l}{\sqrt{\varepsilon} \cdot r_2}$. For each curve of Fig. 2 the condition $r_{2,TE_{011}} = \frac{3.8317 \cdot \lambda_0}{2\pi \cdot \sqrt{\varepsilon}}$ is indicated as a diamond symbol. The line which joins the above points is a guide for the eye, and extrapolates the values given by intermediate permittivity. The shaded region indicates the conditions in which Eq. (2) is not verified or, equivalently, the conditions in which the lowest cutoff frequency of the metallic waveguides is lower than the resonance frequency of the $TE_{011}$ mode. The transformation from the geometry of Fig. 1b to that Fig. 1c is expected to leave this mode confined in the resonator for any point of the allowed region of Fig. 2, at least under the (unphysical) assumption that this transformation doesn't modify its resonance frequency.

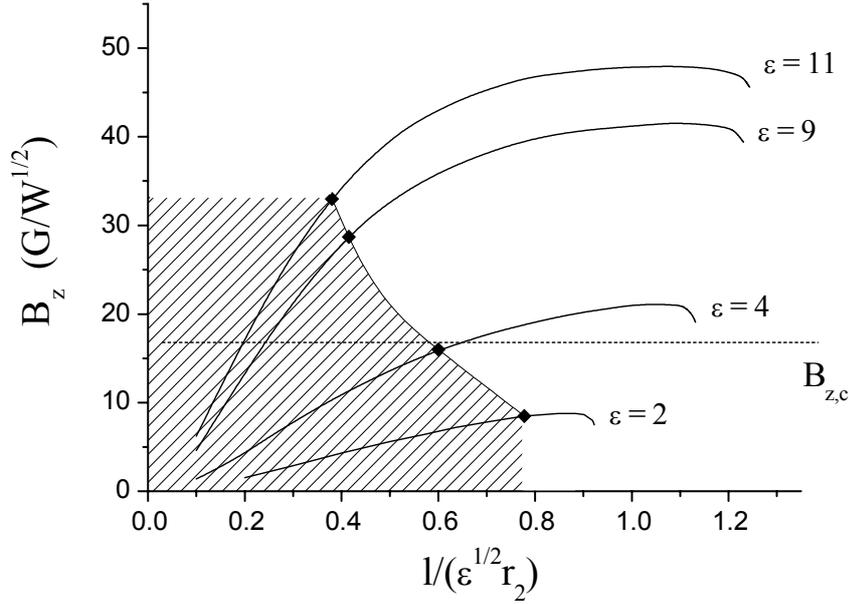

FIG. 2. Chart of the allowed dielectric and geometrical conditions for the $TE_{011}$ mode of an axially open NR resonator. The curves represent the conversion factor $B_z$ vs. the normalized aspect ratio. The shaded area indicates the conditions in which the $TE_{011}$ mode can leak axially. The dashed line represents the conversion factor $B_z$ of the $TE_{011}$ mode of an ideal aluminium cavity, calculated assuming the optimal aspect ratio $l_c=2r_c$.

In a realistic case the validity of the above results can be evaluated by analyzing the resonance frequency variation induced by the axial holes. A possible approach to this problem is based on a perturbative analysis, which gives the frequency variation due to a small deformation of the metallic boundary of the resonator. In particular, the first order variation of the resonance frequency can be written as $\frac{\omega-\omega_0}{\omega_0}=\frac{\int_{\Delta V}\left(\mu\langle H_0^2\rangle-\varepsilon\langle E_0^2\rangle\right)dV}{\int_V\left(\mu\langle H_0^2\rangle+\varepsilon\langle E_0^2\rangle\right)dV}$ [18], where $\Delta V$ is the inward variation of the conducting boundary of the resonator, V is its full volume, and $H_0$ and $E_0$ the unperturbed magnetic and electric field. The brackets represent a time average. A reduction of the volume of the resonator leads to an increase of the resonance frequency when the magnetic energy in the removed volume is predominant. In our case this result can be applied to any intermediate configuration between the initial one, represented by Fig. 1b, and the final one, given by Fig. 1c. The final result will be obviously independent from the intermediate states, which can be chosen in the most convenient way. We will assume intermediate configurations having cylindrical symmetry, in which the dielectric tube finishes with flat end surfaces; the profile of the mirrors will follow the shape of the tube. The corresponding resonance mode, initially given by the $TE_{011}$ mode, remains of $TE_0$ nature, thank to the symmetry of the structure. As a consequence it is characterized by a maximum of magnetic field and a minimum of electric field on the metallic surfaces. For any small elongation of the dielectric tube the resonance frequency decreases, independently of the starting configuration. This allows concluding that the above transformation gives a net reduction of the resonance frequency of the $TE_{011}$ mode. This decrease reinforces the axial NR condition, since reinforces the condition $\nu_{TE_{011}}<\nu_{TE_{01},cutoff}$. As a consequence, the allowed region shown in Fig. 2 is stable with respect to the transformation of a basic NR structure to the axially open one of Fig. 1c. It follows that Fig. 2 can be legitimately assumed as an approximate chart of allowed geometrical and dielectric parameters for the $TE_{011}$ mode. This chart suggests a wide range of proper working resonance conditions for axially open resonators. Wider allowed conditions are expected from a more detailed analysis.

The same inductive demonstration could be in principle developed for the other resonance modes. However, the situation is here complicated by the simultaneous presence of electric and magnetic fields on the metallic boundary of the resonator. Being the other modes by far less important than the $TE_{011}$ one, we will limit now to generic considerations, postponing further comments after the experimental results. In the case of the $TM_{011}$ mode, the lowest cutoff frequency for the $TM_0$ counterpart propagating along the circular waveguide is given by $\nu_{TM_{01},cutoff} = \dfrac{2.4048 \cdot c}{2\pi r_2 \cdot \sqrt{\varepsilon}}$, where 2.4048 represents the first root of the Bessel function $J_0$. The condition $\nu_{TM_{011}} < \nu_{TM_{01},cutoff}$ is now stronger than that found for the $TE_{011}$ mode. Moreover, simulations and measurements reported in Refs. [13, 14] show higher resonance frequencies for the $TM_{011}$ mode than for the $TE_{011}$ one. Accordingly, a substantially reduced region of proper resonant conditions is here expected.

In the case of the $HE_{111}$ mode, the lowest cutoff frequency must be chosen among that ones of both the $TE_{1m}$ and $TM_{1m}$ waveguide modes, due to its hybrid nature. The condition to satisfy is now given by $\nu_{HE_{111}} < \nu_{TE_{11},cutoff} = \dfrac{1.8412 \cdot c}{2\pi r_2 \cdot \sqrt{\varepsilon}}$, which corresponds to a quite low cutoff frequency. On the other hand the measured $HE_{111}$ resonance frequencies are much lower than those of the corresponding $TE_{011}$ mode; no clear indications are then possible at this stage.

The above considerations can be generalized to the case of dielectric tubes having finite length. The thickness of the mirrors will be still assumed much greater than the extension of the evanescent field. Different cases can be considered. If the axial extension of the tube is much beyond the extension of the evanescent field, its length is obviously irrelevant for the resonance mode. If the tube finishes in the region of the evanescent field, a weak effect is expected on the resonance frequencies. On the other hand, the axial confinement of the radiation should be reinforced, due to the increased cutoff frequencies of the empty metallic waveguide. Finally, when the length of the tube is reduced until it finishes inside the mirrors, an abrupt increase of the resonance frequency is expected due to the reduced optical thickness of the resonator.

Having predicted the existence of proper resonant modes in axially open NR structures, the next step concerns a possible excitation scheme. In this respect, the basic similarity between the conventional NR resonators and the axially open one here proposed suggests a direct use of the reflection configuration employed in [13]. A further issue concerns the tuning of the resonance frequency. Two simple approaches can be here envisaged. The first one consists in a reduction of the height of the dielectric tube in the active region of the resonator, as discussed above. The second one is based on the variation of the distance between the mirrors.

The next sections will be devoted to the confirmation of the above considerations. In the interpretation of the obtained results the help of the basic NR configuration of Fig. 1b will be freely invoked.

### 3. Experimental results

The material chosen for the dielectric tube was high-purity, ultra-low OH⁻ contents Suprasil® fused quartz, commercially available from Polymicro Technologies, LLC (Arizona, USA). This choice is first motivated by the relatively high permittivity of fused quartz, $\varepsilon \approx 3.8$, which ensures a large region of allowed working conditions as suggested by the chart of Fig. 2. The dielectric absorption is also acceptable; the loss factor is expected indeed lower than $10^{-3}$ at millimeter wavelengths [21].

The resonance frequency of the $TE_{011}$ mode will be fixed again at 91 GHz. The approximate size of the dielectric region can be defined with the help of Fig. 2. In particular, inside the allowed area the highest conversion factor is expected for a normalized aspect ratio of the order of unity; the related dimensions are given by $r_2 \sim 0.8$ mm and $l \sim 1.5$ mm. The diameter of the dielectric tube was then fixed to $1.60 \pm 0.01$ mm. A central hole of $0.40 \pm 0.01$ mm was foreseen in order to obtain a realistic working configuration.

The planar mirrors were realized by means of two rectangular aluminium plates, each 6 mm thick. A series of 1.7 mm holes were drilled in both plates at corresponding positions. Subsequent holes were positioned at increasing distances from the boundary of the plates in order to allow different coupling ranges. The chosen diameter allows a comfortable insertion of the dielectric tube. The relatively wide tolerance with respect to the diameter of the tube appears appropriate in view of low temperature applications, since it takes into account the different thermal expansion of fused quartz and aluminium. Moreover, the capability of the resonator to allow a significant degree of mechanical tolerance is a fundamental prerequisite for higher frequency applications. The thickness of the aluminium plates was prudentially assumed quite high, since the evanescent decay of the radiation along the holes can extend in principle over long distances.

In the first measurements the distance $l$ between the mirrors was fixed to $1.65 \pm 0.1$ mm. A suitable distance between the dielectric tube and the boundary of the mirror was given by $1.6 \pm 0.1$ mm. The final structure of the employed device is reported in Fig. 3.

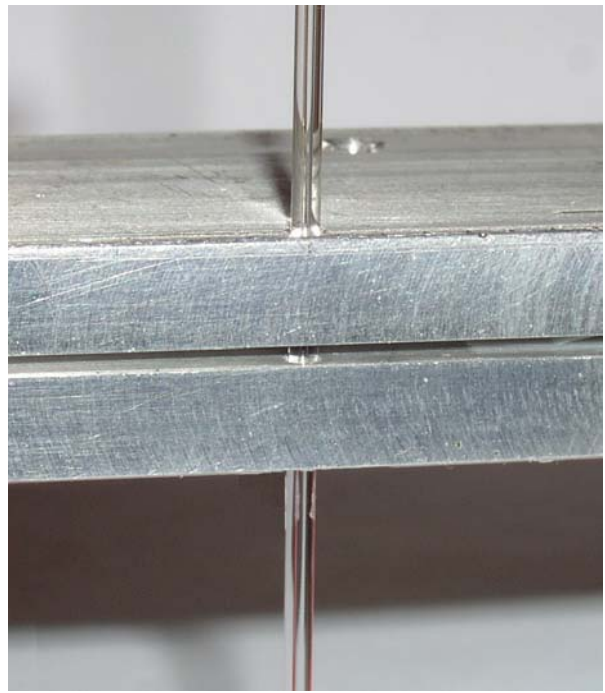

FIG. 3. Picture of the axially open configuration employed in the measurements. The active volume is given by the dielectric region between the two mirrors.

When the above dimensions are assumed for the NR configuration without axial holes, the expected resonance frequencies of the first two modes are given by $\nu_{HE_{111}} = 74.41$ GHz and $\nu_{TE_{011}} = 88.91$ GHz. In these conditions the $TM_{011}$ mode is outside the nonradiative region. The knowledge of these resonance frequencies will give useful information on the effects of the axial holes.

The structure of Fig. 3 was investigated in a frequency interval ranging from 50 to 92 GHz. The coupling of the resonator to the incoming radiation was optimized on the $TE_{011}$ mode. A normalization procedure allowed the extraction of the true spectrum from the signal background. The reflected signal was first acquired in the proper coupling conditions; the resonator was then completely decoupled from the incoming radiation. The reference signal obtained in this latter condition is finally employed to extract the true spectrum of the resonator. Two different techniques were employed to decouple completely the resonator. The first one consists in the movement of the NR device orthogonally to the incoming radiation, as suggested in Sect. 2. The normalized curve obtained in this way is shown in the upper part of Fig. 4. The second option for the acquisition of the signal background is simply to remove the dielectric tube. The result of this latter procedure is shown in the lower part of Fig. 4.

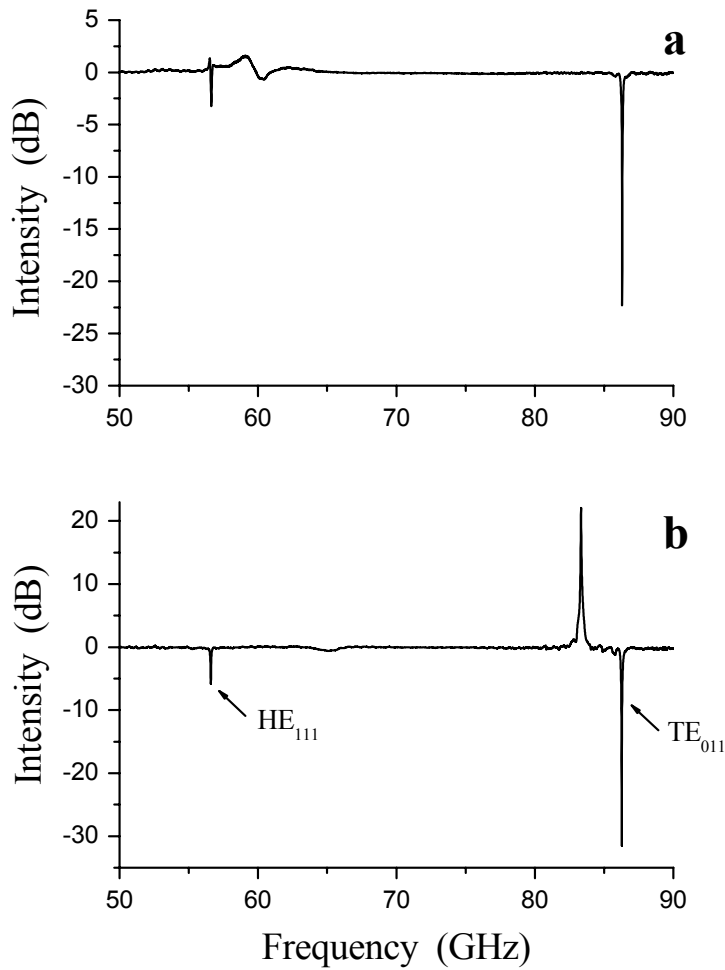

FIG. 4. Normalized spectrum of the resonator of Fig. 3. **a** Normalization obtained displacing the resonating device. **b** Normalization obtained removing the dielectric tube. The assignment of each resonance signal is indicated.

The spectra obtained with the above procedures show some discrepancy. A comparison between them allows however a clear identification of the actual resonances with respect to the artefacts introduced by the normalization procedure [22]. All further normalizations will be done removing the resonator. Two resonance modes can be identified in the curves of Fig. 4. The first one, which is supposed to be the $HE_{111}$ mode, is centered at 56.58 GHz and coupled for 6.1 dB. The corresponding unloaded merit factor is given by $Q_{0,HE_{111}}$=880. The axial holes induce here a strong reduction of the resonance frequency with respect to the corresponding configuration without holes. As a consequence a substantial rearrangement of the field distribution is expected. The second resonance, assigned to the $TE_{011}$ mode, is centered at 86.22 GHz and coupled for 36 dB. The unloaded merit factor is given by $Q_{0,TE_{011}}$=1060. The moderate resonance frequency reduction indicates here a limited distortion of the electromagnetic field.

In order to verify the tuning procedures proposed in the previous section, the axial position of the quartz tube was displaced until one of its ends was at the same level of the lower mirror, as shown in the drawing of Fig. 5.

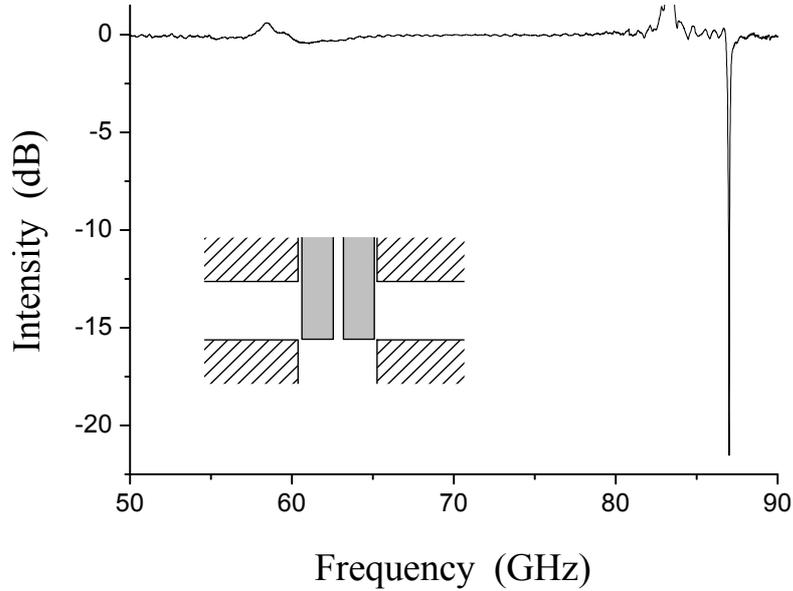

FIG. 5. Normalized spectrum obtained displacing the dielectric tube, as shown in the inset.

The resulting spectrum is given by the curve of Fig. 5. The lower frequency resonance disappears, or at least appears as a quite broad line. The $TE_{011}$ mode is on the other hand still visible, and centered at 86.92 GHz; the tuning of its resonance frequency was about +0.7 GHz. The coupling of the resonance is given by 22 dB and the related merit factor $Q_{0,TE_{011}}$=806. The reduction of this factor can be ascribed to the misalignment of the tube with respect to the axis of the structure. The loss of the axial symmetry generates indeed a mixing among different resonance modes, which open in turn channels for electromagnetic leakage. A similar reason could explain the disappearance of the $HE_{111}$ mode.

In order to test the second tuning mechanism, the distance between the mirrors was reduced to $1.25 \pm 0.1$ mm. The resonance frequencies of the corresponding NR configuration without axial holes are now given by $\nu_{HE_{111}}=88.5$ GHz, $\nu_{TE_{011}}=102.5$ GHz, and $\nu_{TM_{011}} = 113.1$ GHz. A reduction of the distance between the dielectric tube and the boundary of the mirrors was necessary in order to ensure an efficient excitation of the resonator. A lower distance $l$ leads indeed to higher cutoff frequencies for the parallel-plates TE and TM modes, which increase the decay of the evanescent field outside the dielectric region. A proper distance between the tube and the boundary of the mirror was given by $1.1 \pm 0.1$ mm. The spectrum of the resonator obtained in this condition is reported in Fig. 6.

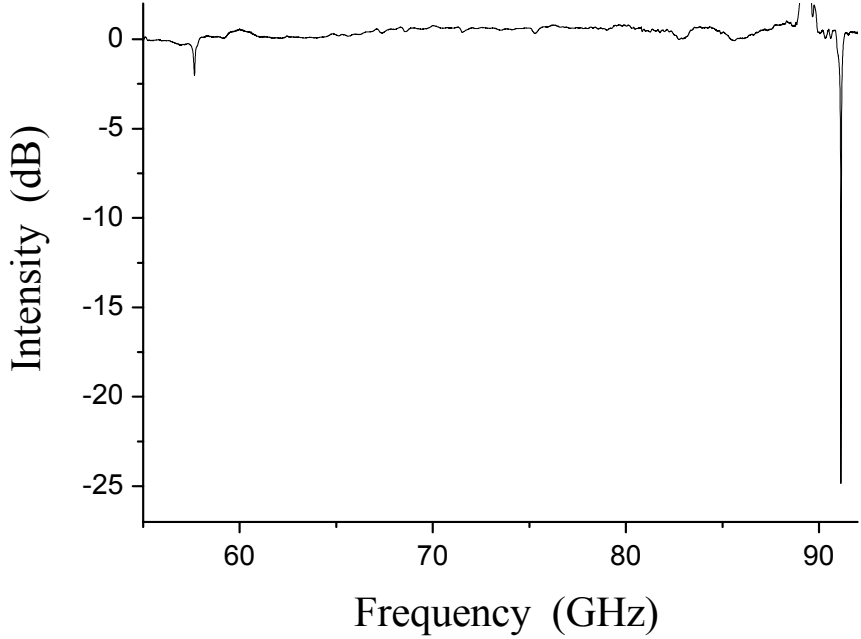

FIG. 6. Normalized spectrum obtained in the configuration of Fig. 3, after a decrease of the distance between the mirrors.

The $HE_{111}$ and the $TE_{011}$ modes were clearly visible again. The $HE_{111}$ mode was centered at 57.66 GHz; the corresponding unloaded merit factor was $Q_{0,HE_{111}}$ =700. The resonance frequency was then tuned by +1.1 GHz. The $TE_{011}$ mode was centered at 91.11 GHz and tuned by about +5 GHz. Its coupling is about 28 dB; the corresponding unloaded merit factor $Q_{0,TE_{011}}$ =1510.

The previous results give indirect information on the field distribution of both modes. In the case of the $HE_{111}$ resonance, the tuning due to the above variation of $l$ is much lower than that induced by the transformation of the structure from a basic NR to an axially open one. This suggests that the rearrangement of the electromagnetic field due to the holes is much higher. The field should then extend much more along the axial holes than outside the dielectric region. In the case of the $TE_{011}$ mode the displacements of the resonance frequency are comparable; the field should then extend comparably in the axial holes and outside the dielectric region.

For ease of comparison, the three $TE_{011}$ modes observed in the investigated configurations are reported in Fig. 7, together with their fit.

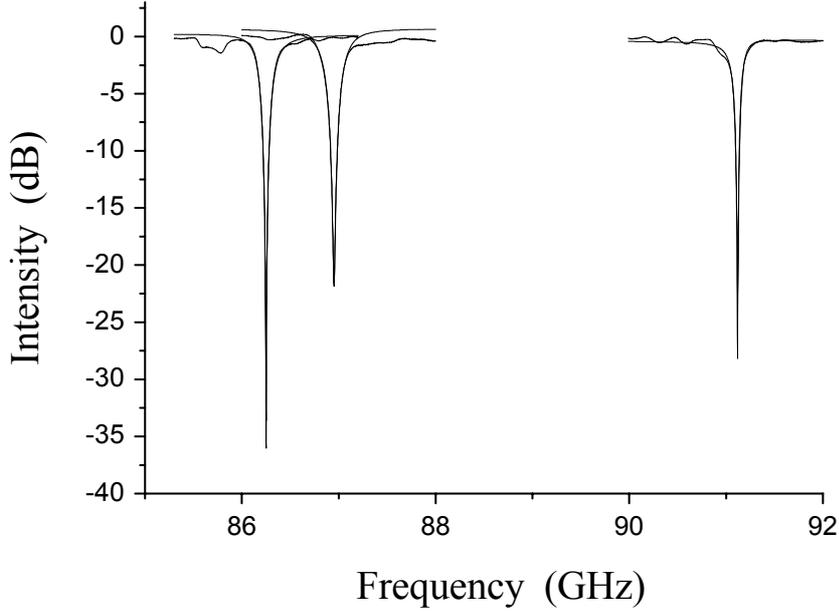

FIG. 7. Detail of the $TE_{011}$ resonances reported in Figs. 4, 5 and 6, together with their lorentzian fit.

In none of the previous measurements there was evidence of the $TM_{011}$ mode, at least in the frequency range of analysis.

## 4. Numerical modelling

Additional information on the observed resonances can be obtained with the aid of computational methods, which can give field distribution, resonance frequency and quality factor for realistic configurations. We will focus in particular on the resonances of Fig. 6, namely for the case of distance between mirrors equal to $1.25 \pm 0.1$ mm. The employed software is the finite element method FEMLAB 3.0a (COMSOL, Sweden). A basic parameter of this numerical approach is represented by the number of elementary regions (mesh elements) in the volume of analysis. Better accuracies are in general obtained for an increasing number of mesh elements. Being this number limited by practical reasons, it is important to reduce as much as possible the volume of analysis. This can be done by exploiting the symmetry of the problem and the properties of the modes of interest. In our case, these modes are given by the $HE_{111}$, the $TE_{011}$ and the $TM_{011}$. The planar symmetry of the problem can be employed to reduce the modelling to the upper half of the resonator. For the above modes the plane of symmetry can be replaced by a magnetic wall [18]. The rotational invariance can be exploited to limit the analysis to a sector of the full structure. In particular, a quarter of the resonator can be used for all the above modes, provided a proper combination of electric and magnetic walls is used for the radial planes [23, 24]. The resulting geometry of analysis is reported in Fig. 8. Here the region of modelling represents an octant of the actual structure. The shaded area indicates the dielectric region.

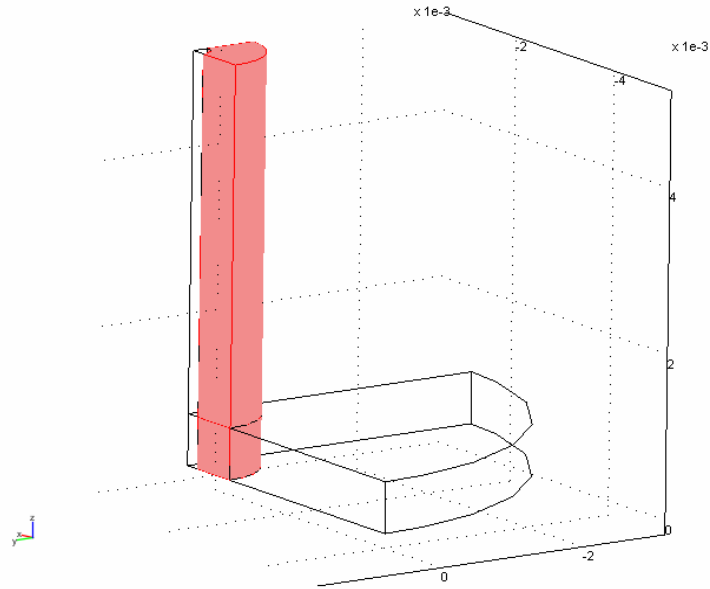

FIG. 8. Section of the axially open NR resonator investigated with the finite element method. The shaded area indicates the dielectric region.

Some approximations were introduced for sake of simplicity. The diameter of the dielectric tube was assumed equal to that of the holes, namely 1.7 mm. All corners were considered perfectly sharp. The height of the metallic mirrors was finally fixed to 5 mm, instead to 6 mm as in the experimental case. A perfect electric wall was put initially at the end of the axial holes. These assumptions are consistent with the qualitative nature of this computational analysis.

The imposition of boundary conditions suitable for the $HE_{111}$ mode gave a first resonance at 54.54 GHz, whose field distribution is reported in Fig. 9. This solution is then compatible with the low frequency resonance experimentally observed, and already indicated as $HE_{111}$ mode. The electromagnetic field penetrates in a substantial way inside the axial holes, in agreement with the analysis of the previous section. The discrepancy with the measured frequency is most probably due to the simplification operated in the modelling. On the contrary, the obtained field distribution is not influenced by the assumption of holes having finite height. The increase of this height from 5 to 6 mm leaves the calculated resonance frequency stable within $10^{-3}$.

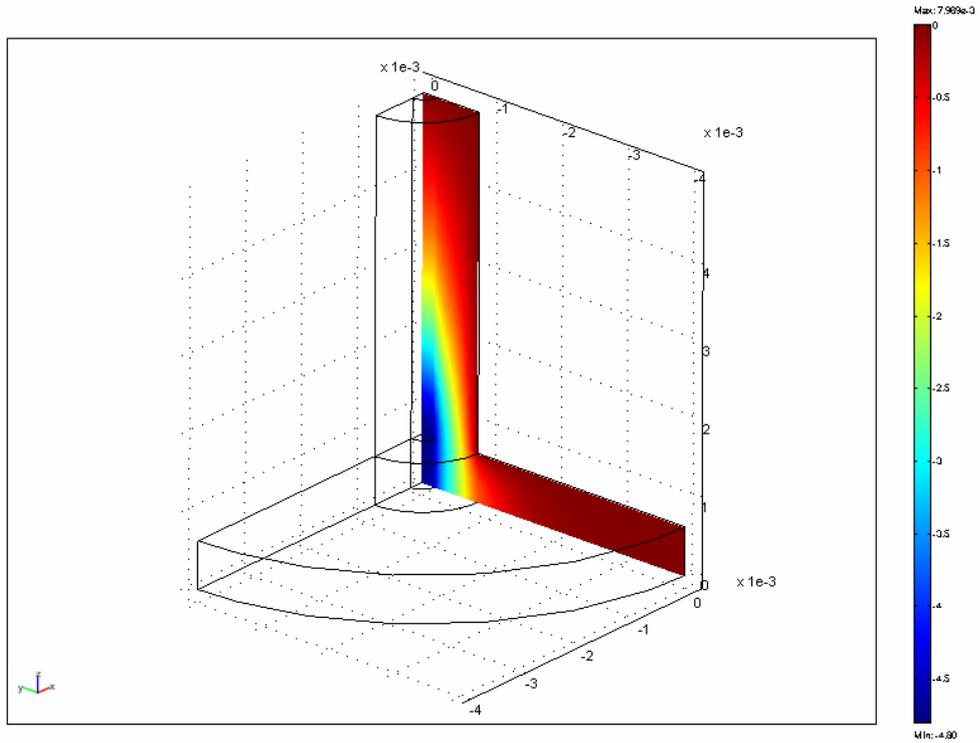

FIG. 9. Computed azimuthal electric field of the $HE_{111}$ mode, together with the geometry of analysis. On the right the colours scales.

The use of boundary conditions compatible with the $TE_{011}$ mode gives a resonance at 91.78 GHz, whose field distribution is reported in Fig. 10. The substantial agreement with the experimental result confirms the assignment of this resonance.

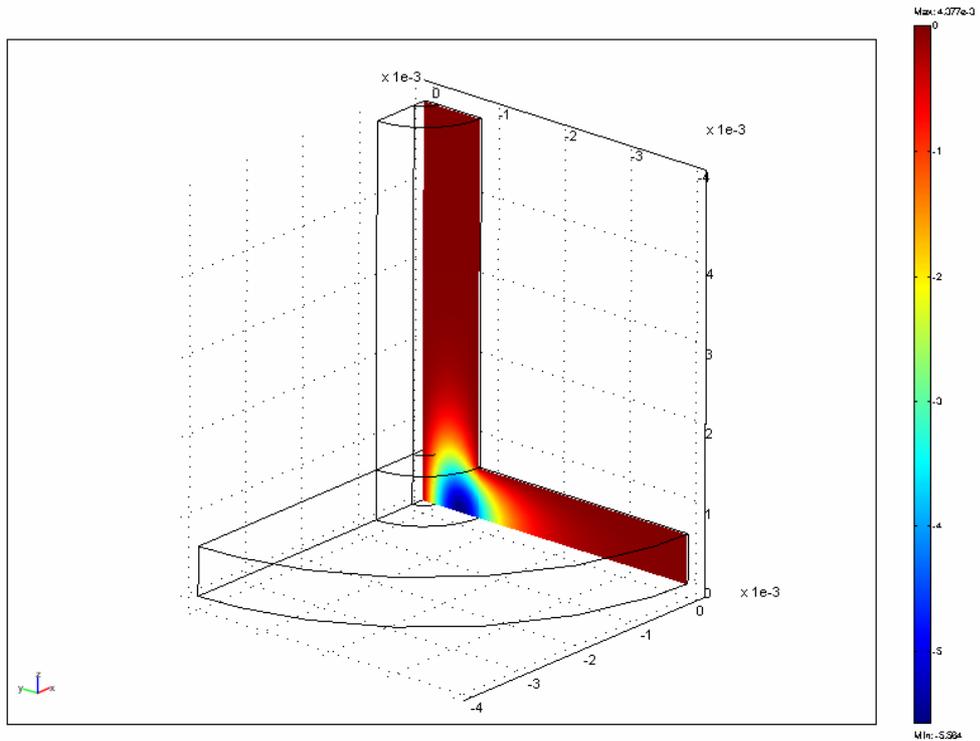

FIG. 10. Computed azimuthal electric field of the $TE_{011}$ mode, together with the geometry of analysis. On the right the colours scale.

In this case the field distribution is essentially limited to the dielectric volume between the conducting mirrors, as suggested in the previous section. The active region of the resonator is then quite small, and similar to that of the corresponding NR configuration without axial holes. The length of the dielectric region and the thickness of the mirrors are now much less critical parameters.

A similar analysis was developed for the $TM_{011}$ resonance by imposing the proper boundary conditions. The modelling showed in this case that, when the profile of the resonator is modified from the basic configuration to the axially open one, the mode follows the boundary inside the holes. As a consequence, for the geometry under analysis the $TM_{011}$ resonance cannot be confined in a bounded region.

The above results can be generalized by using a Perfect Matched Layer (PML), which is a non-physical medium able to absorb almost perfectly the incident radiation [25]. A PML can be used to simulate the behaviour of an open structure in finite modelling volumes. In our case possible leakages of radiation can be monitored including proper PML regions far from the expected active volume of the resonator. The results of this analysis confirm that the $HE_{111}$ and the $TE_{011}$ modes do not irradiate appreciably, while the $TM_{011}$ is distorted and absorbed by the PMLs.

## 5. Discussion and conclusions

The obtained results confirm the basic predictions formulated in Sect. 2. The first two resonance modes of an axially open resonator were observed, namely the hybrid $HE_{111}$ mode and the transverse-electric $TE_{011}$ mode. For the former resonance the experimental and computational results indicate a relevant extension of the fields along the axial holes. This behaviour is expected peculiar of the $HE_{111}$ mode, due to the low cutoff frequency of the corresponding waveguide $TE_{11}$ mode. The $HE_{111}$ mode is then very sensitive to the diameter of the holes and to the thickness of the mirrors. The extension of the mode inside the holes can be limited reducing the length of the dielectric tube, as confirmed by the numerical modelling.

The $TE_{011}$ mode can be well confined in the dielectric region between the metallic mirrors. This reduced extension gives an additional justification to the chart of Fig. 2, obtained under the assumption that the axial holes introduce a weak perturbation. The minimum overall size of the resonator can be reduced in this case up to few wavelengths without appreciable effects on the confinement of the field. The $TE_{011}$ mode is then an ideal candidate for the realization of single-mode resonators having high conversion factor. The performances of the $TE_{011}$ resonance here observed can be evaluated by resorting again to the NR configuration without holes. The height of the disc can be adjusted until the measured resonance frequency is achieved. The value of $l$ corresponding to 91.11 GHz is given by $l = 1.576$ mm. The merit factor of this resonance can be first obtained taking into account the aluminium conductivity only. The value obtained in this way is given by $Q_{0,met} = 5439$. The measured $Q_{0,TE_{011}} = 1510$ can be then reproduced introducing in the modelling a proper dielectric absorption. The loss factor of the Suprasil® tube follows as by-product of this procedure. It results $\tan\delta = 6.7 \cdot 10^{-4}$ around 90 GHz, in agreement with the value given at millimeter wavelengths [21]. Having reproduced the experimental merit factor, the conversion factor can be calculated from the field distribution. The so obtained value, given by $B_z = 18$ G/W$^{1/2}$, represents a reasonable estimation of the conversion factor of the employed resonator. The above procedure is basically equivalent to assume that the active volume of the actual configuration is the same of that of the model configuration. This approximation is supported by the TE$_{011}$ field distribution shown in Fig. 10. The obtained merit factor and conversion factor can be compared with those of an ideal TE$_{011}$ cavity realized with aluminium. In this case it results $Q_{0,c} = 7778$ and $B_{z,c} = 16.77$ G/W$^{1/2}$. Similar values have been experimentally demonstrated in Ref. [5]. Although the quality factor here obtained is substantially lower than that of the ideal metallic cavity, the conversion factor is expected of the same order or better [26]. This is due to the reduced active volume of a dielectric resonator, which in turn leads to enhanced energy densities. The typical volume of a single-mode dielectric resonator, given by $\left(\frac{\lambda_0}{2 \cdot \sqrt{\varepsilon}}\right)^3$, is in particular about $\varepsilon^{3/2}$ times smaller than that of a metallic cavity working at the same frequency.

Despite the extreme ease of realization and the relatively wide mechanical tolerances, the above demonstrated resonator shows state-of-the-art performances at millimeter wavelengths. These performances are mainly limited by the absorption of fused quartz. Better results are expected at lower temperature, where this absorption decreases considerably. The use of better conductors as copper or silver should give an appreciable quality factor improvement as well. The employment of commercial dielectric tubes is one of the specific benefits of the proposed configuration. Being these tubes available on customer specifications down to an external diameter of $150 \pm 6$ μm (minimum internal diameter $2 \pm 1$ μm), the realization of higher frequencies single-mode resonators with well-defined properties can be easily accomplished. The identification between resonator and sample-holder opens new possibilities. Among them, the simple and inexpensive replacement of the resonator and the rotation of the sample without need of *ad hoc* designs. The possibility to characterize different samples inserted along the sample holder can be accounted for as well. In the specific case of EPR spectroscopy also the use of Suprasil® seems strategic, since this material is proven free from appreciable paramagnetic impurities also in high-sensitivity instrumentations [7]. The proposed resonator shows finally all the benefits of a typical NR configuration, as the widely open structure and the presence of few non-degenerate resonance modes.

In conclusion, a single-mode resonator based on a low-loss dielectric tube has been predicted and demonstrated following the concept of axially open NR configuration. Simple excitation and tuning mechanisms have been demonstrated as well. State-of-the-art performances have been obtained at millimeter wavelengths. The proposed open configuration can be easily generalized following the basic conditions governing the behaviour of these rotationally invariant devices, namely $l < \frac{\lambda_0}{2}$ and $r < \frac{\bar{u}_{n,m} \cdot \lambda_0}{2\pi \cdot \sqrt{\varepsilon}}$, where $n$ is the azimuthal modal index of the mode under analysis and $\bar{u}_{n,m}$ a proper root of the first order Bessel function or of its derivative. The obtained results suggest a novel strategy in the design of simple and efficient single-mode resonators for millimeter-wave applications.

**Acknowledgments**

The help of C. A. Massa in the preparation of the experimental setup is gratefully acknowledged.

[24] For the rotationally invariant $TE_{011}$ and $TM_{011}$ modes the volume of analysis can be arbitrarily reduced. The proposed solution is however compatible will all modes of interest, as discussed in the text.

[25] Jianming Jin, *The Finite Element Method in Electromagnetics* (Wiley, New York, 2002)

[26] In specific applications, as pulse magnetic resonance, resonators having high conversion efficiencies for low merit factors represent the most appropriate solution, since they combine high sensitivity with low dead time.